\documentclass[aps,prd,preprintnumbers,superscriptaddress,nofootinbib,%showpacs,
twocolumn]{revtex4-1}%
\usepackage[dvipdfmx]{graphicx}
\usepackage{bm,latexsym,amsmath,amssymb,amsfonts}
\usepackage[dvipdfmx]{color}
\input{colordvi.tex}
\usepackage[dvipdfmx]{hyperref}
\usepackage{url}
\hypersetup{
    colorlinks=true,
    citecolor=cyan,
}
\begin{document}

\title{Parity-violating gravity and GW170817}

\author{Atsushi~Nishizawa}
\email[Email: ]{anishi@kmi.nagoya-u.ac.jp}
\affiliation{Kobayashi-Maskawa Institute for the Origin of Particles and the Universe, Nagoya University, Nagoya 464-8602, Japan}

\author{Tsutomu~Kobayashi}
\email[Email: ]{tsutomu@rikkyo.ac.jp}
\affiliation{Department of Physics, Rikkyo University, Toshima, Tokyo 171-8501, Japan}

\begin{abstract}
  We consider gravitational waves (GWs) in generic parity-violating gravity
  including recently proposed ghost-free theories with parity violation
  as well as Chern-Simons (CS) modified gravity, and study the implications of observational constraints from GW170817/GRB 170817A.
  Whereas GWs propagate at the speed of light, $c$,
  in CS gravity,
  we point out that this is specific to CS gravity and
  the GW propagation speed deviates from $c$,
  in general, in parity-violating gravity.
Therefore, contrary to the previous literature in which
  only CS gravity is studied as a concrete example, we show that
  GW170817/GRB 170817A can, in fact, be used to limit gravitational parity violation.
Our argument implies that the constraint on the propagation speed of GWs can
pin down the parity-violating sector, if any, to CS gravity.
\end{abstract}

\preprint{RUP-18-29}
\maketitle
%%%%%%%%%%%%%%%%%%%%%%%%%%%%%%%%%%%%%%%%%%%%%%%%%%%%%%%%%%%%%%%%%%%%%%
\section{Introduction}

The first detection of gravitational waves (GWs)
from two merging black holes, GW150914~\cite{GW150914:detection}, has opened a
new and intriguing arena for gravitational physics.
Direct observation of GWs provides us a window into
the regime of strong gravity and the propagating sector of gravity,
enabling novel tests of general relativity.
More recently, the nearly simultaneous detection of GWs and
the gamma-ray burst from the merger of neutron stars,
GW170817/GRB 170817A~\cite{GW170817:detection,GW170817:GRB}, gave us an unprecedented opportunity to measure the speed of GWs, $c_{\rm T}$ at a level of one part in $10^{15}$. 

The physics of propagation of GWs is simple and clear compared to that of generation.
Before the occurrence of GW170817/GRB170817A \cite{GW170817:detection}, it had been expected that comparing arrival times between GW from the merger of neutron stars and high-energy photons from a short gamma-ray burst would allow us to measure GW speed so precisely \cite{Nishizawa:2014zna} and consequently tightly constrain the modification of gravity relevant to the cosmic accelerating expansion \cite{Lombriser:2015sxa}. Indeed the speed bound from GW170817/GRB 170817A
constrains modified gravity theories at the precision of $10^{-15}$ and a large class of theories as alternatives to dark energy have already been almost ruled out~\cite{Baker:2017hug,Creminelli:2017sry,Sakstein:2017xjx,Ezquiaga:2017ekz,Arai:2017hxj,Gumrukcuoglu:2017ijh,Oost:2018tcv,Gong:2018cgj,Gong:2018vbo} (see, however,~\cite{deRham:2018red}).

%Nishizawa:2016kba

In this paper we further pursue the implications
of GW170817/GRB 170817A for modified gravity.
Earlier works~\cite{Baker:2017hug,Creminelli:2017sry,%
Sakstein:2017xjx,Ezquiaga:2017ekz,Arai:2017hxj,%
Gumrukcuoglu:2017ijh,Oost:2018tcv,Gong:2018cgj,Gong:2018vbo}
focus only on parity-preserving theories.
Based on one specific parity-violating realization
called Chern-Simons (CS)
gravity~\cite{Jackiw:2003pm} (see also~\cite{Lue:1998mq}),
it is argued that the constraint on $c_{\rm T}$
places no bounds on gravitational parity violation~\cite{Alexander:2017jmt}.
In this paper, we revisit this point.

As stated above, gravitational parity violation has been studied
mostly through CS gravity as a concrete example,
whereas recent developments in modified gravity have revealed that, in fact,
one can construct theories of parity-violating gravity
other than CS gravity~\cite{Crisostomi:2017ugk}.
This leads us to consider a unifying framework to study
the propagation of GWs in generic parity-violating gravity.
We clarify how special CS gravity is among parity-violating theories,
and
show that the bound on the speed of GWs yields a stringent
constraint on parity violation in theories other than CS gravity.

%%%%%%%%%%%%%%%%%%%%%%%%%%%
\section{Parity-violating gravity}

We consider parity-violating gravity whose action is of the form
\begin{align}
S=\frac{1}{16\pi G}\int d^4x\sqrt{-g}\left[
R+{\cal L}_{\rm PV}+{\cal L}_\phi
\right],\label{covariant_action}
\end{align}
where $R$ is the Ricci scalar, ${\cal L}_{\rm PV}$ is a
parity-violating Lagrangian, and ${\cal L}_\phi$ is the Lagrangian
for a scalar field $\phi$ which may be coupled nonminimally to gravity.

The most frequently studied example of parity-violating gravity
is CS gravity for which ${\cal L}_{\rm PV}$ is given by
\begin{align}
{\cal L}_{\rm PV}={\cal L}_{\rm CS}:= f(\phi) P,\quad P:=
\varepsilon^{\mu\nu\rho\sigma}R_{\rho\sigma\alpha\beta}
R^{\alpha\beta}_{~~\;\mu\nu},
\end{align}
where $\varepsilon^{\mu\nu\rho\sigma}$ is the Levi-Civit\`{a} tensor
defined as $\varepsilon^{\mu\nu\rho\sigma}:=\epsilon^{\mu\nu\rho\sigma}/\sqrt{-g}$
with $\epsilon^{\mu\nu\rho\sigma}$ being the antisymmetric symbol.
The Pontryagin term $P$ is a topological invariant in four dimensions,
and for this reason we need the dynamical scalar field $\phi$
coupled to $P$ via $f(\phi)$
in order for this term to contribute to the field equations.
The kinetic term for $\phi$ is supposed to be included in ${\cal L}_\phi$
and one of the simplest possibilities is
${\cal L}_\phi=-(\partial\phi)^2/2-V(\phi)$.

Since CS gravity has higher-derivative field equations
as explicitly confirmed by varying the action with respect to
the metric~\cite{Jackiw:2003pm,Alexander:2009tp},
one expects that dangerous Ostrogradsky ghosts appear in this theory.
This is indeed true, as can be seen directly, e.g., from
a wrong sign kinetic term in the quadratic action for perturbations around
a spherically symmetric background~\cite{Motohashi:2011ds}
(see also~\cite{Dyda:2012rj}).
This conclusion is supported by the Hamiltonian analysis performed in~\cite{Crisostomi:2017ugk}.
The ghost degrees of freedom might not be problematic if
the theory (\ref{covariant_action}) is treated as a low-energy
truncation of a fundamental theory, but they do cause
instabilities if regarded as a complete theory. Note that at least in the unitary gauge in which $\phi$ is homogeneous on $t=\;$const hypersurfaces, CS gravity is ghost-free~\cite{Crisostomi:2017ugk}.

Recently, ghost-free parity-violating theories of gravity
have been explored in~\cite{Crisostomi:2017ugk}. At least in the unitary gauge,
it is found that one can indeed construct Ostrogradsky-stable theories other than CS gravity.
One of the theories proposed in~\cite{Crisostomi:2017ugk} is given by the
following Lagrangian:
\begin{align}
{\cal L}_{\rm PV1}=\sum_{A=1}^4 a_A(\phi,\phi_\mu\phi^\mu)L_A,
\label{gfpv1}
\end{align}
with
\begin{align}
L_1&:=\varepsilon^{\mu\nu\alpha\beta}R_{\alpha\beta\rho\sigma}
R_{\mu\nu~\lambda}^{~~\;\rho}\phi^\sigma\phi^\lambda,
\\
L_2&:=\varepsilon^{\mu\nu\alpha\beta}R_{\alpha\beta\rho\sigma}
R_{\mu\lambda}^{~~\;\rho\sigma}\phi_\nu\phi^\lambda,
\\
L_3&:=\varepsilon^{\mu\nu\alpha\beta}R_{\alpha\beta\rho\sigma}
R^\sigma_{~\;\nu}\phi^\rho\phi_\mu,
\\
L_4&:=\phi_\lambda\phi^\lambda P,
\end{align}
where $\phi_\mu:=\nabla_\mu\phi$.
In order to remove the Ostrogradsky modes,
it is required that $4a_1+2a_2+a_3+8a_4=0$.
Similarly, another ghost-free, parity-violating theory
found in~\cite{Crisostomi:2017ugk}
contains second derivatives of the scalar field, $\phi^\mu_\nu:=\nabla^\mu\nabla_\nu\phi$,
and
is described by the Lagrangian of the form
\begin{align}
{\cal L}_{{\rm PV}2}=b_1(\phi,\phi_\lambda\phi^\lambda)
\varepsilon^{\mu\nu\alpha\beta}R_{\alpha\beta\rho\sigma}\phi^\rho \phi_\mu
\phi^\sigma_\nu+\cdots,
\end{align}
though its explicit expression is not important in this paper.

The purpose of the present paper is to study the propagation of
GWs in such parity-violating theories of gravity.
Let us consider GWs propagating on
a homogeneous and isotropic background. The spatial metric is written as
$g_{ij}=a^2(t)[\delta_{ij}+h_{ij}(t,\Vec{x})]$ with the scale factor $a(t)$.
To derive the evolution equation for $h_{ij}$, we substitute the
perturbed metric to Eq.~(\ref{covariant_action}) and expand it to
second order in $h_{ij}$.
Let us focus on ${\cal L}_{\rm PV1}$ for the moment.
After some manipulation, we find\footnote{Hereafter we do not discriminate the upper and lower spatial indices because they are interchanged by $\delta_{ij}$ and $\delta^{ij}$.}
\begin{align}
  {\cal L}_{{\rm PV}1}^{(2)}\supset
  \epsilon^{ijk}\dot h_{il}\partial_j\dot h_{kl}
  ,\quad \epsilon^{ijk}\partial^2 h_{il}\partial_j\dot h_{kl},
\label{2terms}
\end{align}
where $\epsilon^{ijk}$ is the antisymmetric symbol and
the coefficients of these terms depends on time in general.
In the language of the Arnowitt-Deser-Misner (ADM) formalism,
the two terms come, respectively,
from $\epsilon^{ijk}K_{il}D_jK_k^l$ and $\epsilon^{ijk}R_{il}^{(3)}D_jK_k^l$,
where $K_{ij}$ and $R_{ij}^{(3)}$ are
the extrinsic and intrinsic curvature tensors of the spatial hypersurfaces
and $D_i$ is the three-dimensional covariant derivative.
The second term in Eq.~(\ref{2terms}) can be recast in
$\epsilon^{ijk}\partial^2 h_{il}\partial_j h_{kl}$
by performing integration by parts.
Therefore, the final form of the quadratic action for $h_{ij}$
is of the form
\begin{align}
S^{(2)}=\frac{1}{16\pi G}\int dt d^3x\,
a^3 \left[{\cal L}_{\rm GR}^{(2)}+{\cal L}_{\rm PV}^{(2)}\right],
\label{action2}
\end{align}
where
\begin{align}
{\cal L}_{\rm GR}^{(2)}=\frac{1}{4}\left[
\dot h_{ij}^2-a^{-2}(\partial_k h_{ij})^2\right]
\label{grlag}
\end{align}
is the standard Lagrangian obtained from
the Einstein-Hilbert term $R$ and
\begin{align}
{\cal L}_{\rm PV}^{(2)}=
\frac{1}{4}\left[
\frac{\alpha(t)}{a\Lambda}\epsilon^{ijk}\dot h_{il}
\partial_j \dot h_{kl}+\frac{\beta(t)}{a^3\Lambda}\epsilon^{ijk}\partial^2 h_{il}
\partial_j h_{kl}\right]
\label{pvlag}
\end{align}
is the Lagrangian signaling parity violation.
Here $\alpha$ and $\beta$ are dimensionless functions of time
and $\Lambda$ is some energy scale.
Note that $\alpha$ and $\beta$ are independent in general. At least either of $\alpha$ and $\beta$ is taken to be an ${\cal O}(1)$ quantity by rescaling $\Lambda$.
We obtain only the first term in Eq.~(\ref{2terms}) if we
start from ${\cal L}_{{\rm PV}2}$.
Therefore, the Lagrangian~(\ref{pvlag}) contains ${\cal L}_{{\rm PV}2}$ as
the special case with $\beta(t)=0$.
By expanding ${\cal L}_{{\rm CS}}$ one sees that
CS gravity corresponds to the special case of
the above general action satisfying
\begin{align}
\alpha(t) = \beta(t).\label{alpha=beta}
\end{align}
(See, e.g.,~\cite{Alexander:2004wk}.)
Thus, the quadratic action~(\ref{action2})
with~(\ref{grlag}) and~(\ref{pvlag}) offers us
a unifying framework to study the propagation of GWs
in parity violating theories of gravity described above.
The concrete forms of $\alpha(t)$ and $\beta(t)$
depend on the background cosmological evolution of $a(t)$ and $\phi(t)$
as well as the theory under consideration.

Note that $\alpha(t)=\beta(t)$ could in principle occur
even in the ${\cal L}_{\rm PV1}$ theory. However, an extreme fine-tuning
of the time-dependent functions is required in the ${\cal L}_{\rm PV1}$ theory,
while Eq.~(\ref{alpha=beta}) is automatically satisfied in CS gravity.

The Lagrangian for CS gravity is sometimes expressed
using some length scale $\ell_{\rm CS}$ and
the dimensionless scalar field $\vartheta$ as
${\cal L}_{\rm CS} \sim (\ell_{\rm CS}^2\vartheta/4) P$,
and $\ell_{\rm CS}$ is often denoted as $\xi^{1/4}$.
This notation can be converted to ours as
$\alpha/\Lambda = \beta/\Lambda \sim \ell_{\rm CS}^2\dot\vartheta/4$
(ignoring the cosmic expansion).

%{(If the scale factor is set to $a=1$ and the Hubble terms are ignored, the correspondence $\alpha/\Lambda = \beta/\Lambda = \ell_{\rm CS}^2\dot\vartheta/4$ holds except for the overall factor of $-16\pi G$. Also the correspondence of $\ell_{\rm CS}^2 = \alpha_{\rm Yagi}$.)}

Before proceeding, let us give two comments on
possible further generalization of the above framework.
The first comment is that
one can generalize the standard piece ${\cal L}_{\rm GR}^{(2)}$
to
\begin{align}
{\cal L}_{\rm H}^{(2)}=\frac{1}{4}\left[
A(t)\dot h_{ij}^2-a^{-2}B(t)(\partial_k h_{ij})^2\right]\label{hormod}
\end{align}
by considering, e.g., the Horndeski/generalized Galileon Lagrangian
as ${\cal L}_\phi$~\cite{Horndeski:1974wa,Deffayet:2011gz,Kobayashi:2011nu}. However, for the moment we assume
the standard Lagrangian~(\ref{grlag}) for the parity-preserving part.

The second comment is that Eq.~(\ref{pvlag}) is not only the unifying description of
the known parity-violating terms
${\cal L}_{{\rm CS}}$, ${\cal L}_{{\rm PV}1}$, and ${\cal L}_{{\rm PV}2}$,
but also the low-energy effective description
of generic parity-violating GWs.
Indeed, the same quadratic action was derived from the viewpoint of the
effective field theory in~\cite{Creminelli:2014wna}.\footnote{In this paper,
the second term in~(\ref{pvlag}) comes from $\epsilon^{ijk}R_{il}^{(3)}D_jK_k^l$
and integration by parts, while in~\cite{Creminelli:2014wna} the same
term is derived directly from the three-dimensional CS term.}
In light of this viewpoint,
one may, for example, further add terms like
$\Lambda^{-5}\epsilon^{ijk}\dot h_{il}\partial^2\partial_j \dot h_{kl}$,
$\Lambda^{-5}\epsilon^{ijk}\partial^2 h_{il}\partial^2\partial_j h_{kl}$,
$\cdots$, which are suppressed by powers of $1/\Lambda$.
The latter was studied in the context of
Ho\v{r}ava gravity~\cite{Takahashi:2009wc}.

%%%%%%%%%%%%%%%%%%%%%%%%%%%
\section{Propagation of parity-violating GWs}

Varying the action~(\ref{action2}) with respect to $h_{ij}$,
we obtain the equation of motion for GWs,
% and expressing the time derivatives in conformal time,
\begin{align}
&h_{ij}''+2{\cal H}h_{ij}'-\partial^2 h_{ij}
\notag \\
&+\frac{1}{a\Lambda}\epsilon_{ilk}\partial_l
\left[\alpha h_{jk}''+
({\cal H}\alpha+\alpha') h_{jk}'
-\beta\partial^2h_{jk}
\right]=0,
%&\left( h^{\prime\prime} + 2 {\cal H} h^{\prime} - \partial^2 h \right)^{ij} \nonumber \\
%& +\frac{1}{2a \Lambda} \epsilon^{i \ell k} \left\{ ({\cal H} \alpha + \alpha^{\prime}) \partial_{\ell} h^{\prime} +\alpha \partial_{\ell} h^{\prime\prime} - \beta \partial^2 \partial_{\ell}h \right\}^{kj} =0 \;,
\label{propagation-eq1}
\end{align}
where the prime denotes differentiation with respect to the conformal time
defined by $d\eta = a^{-1}dt$,
and ${\cal H}:=a'/a$.
%where the prime is the time derivative with respect to conformal time and ${\cal H} \equiv a^{\prime}/a$. The indices, $i,j,k$, are fixed to the GW amplitude $h$.

We decompose $h_{ij}$ into
the circular polarization basis defined by the following linear combination of the
standard $+$ and $\times$ polarization basis:
\begin{align}
e_{ij}^{\rm R} := \frac{1}{\sqrt{2}} (e_{ij}^{+} + i e_{ij}^{\times}) \;, \quad
e_{ij}^{\rm L} := \frac{1}{\sqrt{2}} (e_{ij}^{+} - i e_{ij}^{\times}) \;.
\end{align}%as the linear combinations of $+$ and $\times$ polarizations.
This choice of the polarization basis
is convenient for parity-violating GWs because
the equations of motion for the left and right circular polarizations
are decoupled even though
the parity-violating terms mix the $+$ and $\times$ polarizations.
%but it is diagonalized for the circular polarizations.
Performing a Fourier decomposition, we write
\begin{equation}
h_{ij} (\eta,\vec{x}) = \frac{1}{(2\pi)^{3/2}} \sum_{A={\rm R,L}} \int d^3
 k\, h_{\Vec{k}}^A (\eta) e_{ij}^A e^{i \vec{k}\cdot \vec{x}} \;.
\end{equation}
Using the identity
\begin{equation}
\epsilon_{ilk} n_{l} e_{jk}^{\rm R,L} = i \lambda_{\rm R,L} e^{{\rm R,L}}_{ij} \;,
\label{eq:id}
\end{equation}
where
$\lambda_{\rm R}=+1$, $\lambda_{\rm L}=-1$, and $n_{l}$ is
a unit vector pointing to the direction of propagation, we obtain
\begin{align}
&\left( 1- \lambda_A \tilde{k} \alpha \right) ( h_{\Vec{k}}^A )^{\prime \prime}
 + \left[ 2 - \lambda_A \tilde{k} \left(\alpha + \alpha^{\prime}{\cal H}^{-1}\right) \right]
 {\cal H} ( h_{\Vec{k}}^A )^{\prime}
\nonumber \\
&+ \left( 1-  \lambda_A \tilde{k} \beta \right) k^2 h_{\Vec{k}}^A =0,\label{pvpreq1}
\end{align}
for $A={\rm R}$ and ${\rm L}$.
Here we defined the dimensionless wave number
$\tilde{k} := k/(a\Lambda)$, which controls the magnitude of the corrections to
general relativity. This parameter depends on the frequency of GWs, indicating that
the parity-violating effect is more efficient at higher frequencies such
as LIGO's observation band than
at lower frequencies corresponding, e.g., to
CMB scales. In order to
compare the propagation equation derived above
with the general framework of GW propagation~\cite{Nishizawa:2017nef},
we rewrite Eq.~(\ref{pvpreq1}) in
a physically more transparent form as
\begin{equation}
( h_{\Vec{k}}^A )^{\prime \prime} + (2+\nu_A) {\cal H} ( h_{\Vec{k}}^A )^{\prime}
+ (c_{\rm T}^A)^2 k^2 h_{\Vec{k}}^A =0 \;,
\end{equation}
with the additional amplitude damping and the GW propagation speed squared,
\begin{equation}
\nu_A =
\frac{ \lambda_A \tilde{k}(\alpha-\alpha^{\prime} {\cal H}^{-1})}{1-\lambda_A \tilde{k} \alpha} \;,
 \quad (c_{\rm T}^A )^2 = \frac{1-\lambda_A \tilde{k} \beta }{1-\lambda_A \tilde{k} \alpha } \;.
\label{observables:exact}
\end{equation}

If one takes $\alpha=\beta$, CS gravity is recovered,
in which case we exactly have
$c_{\rm T}^A=1$ and only the amplitude is modified through $\nu_A$.
This agrees with the previous argument~\cite{Yunes:2010yf,Alexander:2017jmt}.
In general parity-violating gravity, however,
the propagation speed is also modified.
Note that since
the sign of $\nu_A$ is determined by $\lambda_A$,
$\nu_{\rm R}$ and $\nu_{\rm L}$
always have opposite signs.
That is, if the amplitude of one polarization mode is enhanced,
the other is suppressed.
This is also true for $(c_{\rm T}^A )^2 -1$:
If one polarization mode is superluminal, then the other is subluminal.

%%%%%%%%%%%%%%%%%%%%%%%%%%%
\section{Observational constraints}

%If it is assumed
Assuming
that the parity-violating effect is a small correction to general relativity, namely,
$\tilde{k} \ll 1$,
the observables given in Eq.~(\ref{observables:exact})
%with the leading corrections
are
\begin{align}
\nu_A &=\lambda_A \tilde{k}(\alpha  -\alpha^{\prime} {\cal H}^{-1})
+{\cal O}(\tilde{k}^2)
 \;, \\
 \left(c_{\rm T}^A \right)^2 &= 1+ \lambda_A \tilde{k} (\alpha-\beta)
 +{\cal O}(\tilde{k}^2)
  \;.
\label{observables:approx}
\end{align}
The GW speed has already been measured from the coincident detections of
GW170817/GRB 170817A~\cite{GW170817:detection,GW170817:GRB} and it is
constrained so tightly in the range $-7 \times 10^{-16} < 1-c_{\rm T} < 3 \times 10^{-15}$.
From this and Eq.~(\ref{observables:approx}),
we have the constraint on parity violating gravity,
$\tilde{k} \left| \alpha-\beta \right| \lesssim 10^{-15}$.
Since the LIGO constraint is for the GW speed at a frequency of $k/a\sim 100\,$Hz,
this can also be written as
\begin{align}
\Lambda^{-1}\left| \alpha-\beta \right| \lesssim 10^{-11}\,{\rm km}.
\label{eq:obs-constraint}
\end{align}
This implies that either of the following statements holds in the low-redshift Universe:
(i) the parity-violating sector is given by CS gravity;
(ii) the parity-violating sector is given by a linear combination of
${\cal L}_{\rm PV1}$ and ${\cal L}_{\rm PV2}$ with $\alpha-\beta={\cal O}(1)$, and
the stringent constraint is obtained as $\Lambda^{-1}\lesssim 10^{-11}\,$km;
(iii) the parity violating sector is given by a linear combination of
${\cal L}_{\rm PV1}$ and ${\cal L}_{\rm PV2}$, and the two time-dependent functions are extremely fine-tuned. Note that in the third case at least ${\cal L}_{\rm PV1}$ is necessary
because ${\cal L}_{\rm PV2}$ generates only the $\alpha$ term.
Note also that, of course, one has the freedom to add ${\cal L}_{\rm CS}$
in the second and third cases.

%The bound on the length scale associated to gravitational parity violation has been obtained for CS gravity. For example, @@@@@@@ $\Lambda^{-1}<????\,$km @@@@@. Our result shows that, if the parity-violating sector is {\em not} given by CS gravity, the bound on this characteristic length scale becomes much tighter thanks to the LIGO constraint, though one should note that various observational constraints should be reconsidered in parity-violating theories other than CS gravity.

So far we have assumed that the parity-preserving part is
described by general relativity.
If one generalizes this part to the Horndeski-type Lagrangian~(\ref{hormod}),
$\nu_A$ and $c_{\rm T}^A$ are also affected by this modification.
However, the two effects cannot be canceled out because
the parity-violating modification depends on
$\lambda_A$ and the wave number, while the Horndeski-type modification does not.
The two ways of modifying gravity are thus distinct, and
therefore the above statements are robust.

% Whereas the stringent constraint derived in this paper is for parity-violating theories other than CS gravity
The above constraint should be taken with a caution when one
compares it with the existing constraints for the typical length scale $\Lambda^{-1}$ in literature. The analysis is made basically for CS gravity only, and
the constraints are
based on the assumptions that
the scalar sector is given simply by ${\cal L}_\phi=-(\partial\phi)^2/2-V(\phi)$
and that $\phi$ has a spacelike gradient, or $\phi$ is considered to be a nondynamical field having the fixed configuration
$f(\phi)\propto \phi ={\rm const}\times t$.
In the latter case, the bound on the typical length scale is
given for example by
$\Lambda^{-1} \lesssim 10^{-1}\,$km~\cite{AliHaimoud:2011bk}, which comes from the double-binary-pulsar observation. This bound can be slightly improved by measuring a GW propagation effect with the future GW detectors \cite{Yagi:2017zhb}.
However, these would depend on the
form of ${\cal L}_\phi$ as well as the parity-violating sector of gravity.
%This fact makes difficult to compare the existing
%constraints with ours straightforwardly.
It should be emphasized that our constraint has been derived without assuming any specific form of the scalar-field Lagrangian.

The observational bound~(\ref{eq:obs-constraint}) constrains
the combination $\Lambda^{-1}\left| \alpha-\beta \right|$ and we cannot distinguish the two possibilities (ii) and (iii) above.
That is, if $\alpha$ and $\beta$ are extremely fine-tuned,
the constraint tells nothing about the energy scale $\Lambda$ of parity violation.
Let us remark that another constraint can possibly come from the measurement
of the gravitational constant at a binary pulsar.
In modified gravity, the effective gravitational constant
for the tensor modes, $G_{\rm GW}$, can be different from that of Newtonian gravity, $G_N$.
However, the binary-pulsar constraint from PSR B1913+16
leads to $0.995 \lesssim G_{\rm GW}/G_N \lesssim 1.00$~\cite{Jimenez:2015bwa}.
This strongly limits, for instance, viable scalar-tensor
theories satisfying $c_{\rm GW}=1$~\cite{Dima:2017pwp}. In our cases,
from the Lagrangians~(\ref{grlag}) and~(\ref{pvlag})
with $\tilde{k} \alpha$ and $\tilde{k} \beta$ now being the same at the level of $10^{-15}$,
the effective gravitational constant for the tensor modes is defined by
including the additional contribution from the parity-violating terms as
\begin{align}
G_{\rm GW}^{A}=G\left(1-\lambda_A \tilde{k} \alpha \right)^{-1}.
\end{align}
Even in the presence of the parity-violating terms, spherically
symmetric solutions remain the same as in general relativity~\cite{Jackiw:2003pm}
if $\phi$ is minimally coupled to matter and gravity (except for
the gravitational parity-violating part). In this case, we may set $G_N=G$.
Then, the binary-pulsar constraint is translated to
$|\tilde{k} \alpha| \lesssim 5 \times 10^{-3}$,
or
\begin{equation}
\Lambda^{-1} |\alpha| \lesssim  10^{6}\,{\rm km} \;,
\end{equation}
using the GW frequency of $k/a\sim 4\times 10^{-4}\,$Hz \cite{Weisberg:2010zz}.
This indicates that the deviation from general relativity due to the parity violation effect must be smaller than $0.5\%$ in the Lagrangian (\ref{action2}).

%\textcolor{blue}{@@@@@@Indeed, the post-Newtonian parameters $\gamma_{\rm PPN}$ and $\beta_{\rm PPN}$ measured in the Solar system coincide well with those in general relativity at the precision of $10^{-5}$ and $10^{-4}$, respectively@@@@@@} \cite{Will:2014kxa}.

%\begin{align}
%L_{\rm GW}\sim \frac{a^3}{64\pi G_{\rm GW} }\left[
%\dot h_{ij}^2-a^{-2}(\partial_kh_{ij})^2
%\right].
%\end{align}

%%%%%%%%%%%%%%%%%%%%%%%%%%%
\section{Conclusions}

We have studied the propagation of gravitational waves (GWs) in
Chern-Simons (CS) modified gravity and recently proposed
ghost-free theories of parity-violating gravity.
Along with this latter extension of gravity, we have found that
the propagation speed of GWs is modified in general, together with
the amplitude damping, which is modified in CS gravity as well.
From the measurement of the GW speed with GW170817/GRB 170817A,
we conclude that the possible parity-violating extension of gravity
at low redshifts has already been tightly restricted to fine-tuned models or CS gravity.
We have not considered any specific Lagrangian for the scalar degree of freedom, ${\cal L}_\phi$.
Our result thus relies only on the propagation of GWs and the assumption that the scalar field $\phi$ has a timelike gradient,
and hence is robust irrespective of this scalar sector (provided that such a scalar-field configuration is allowed).

%%%%%%%%%%%%%%%%%%%%%%%%%%%
\begin{acknowledgments}
A.N. was supported by JSPS KAKENHI Grants No.~JP17H06358 and No.~JP18H04581.
T.K. was supported by
JSPS KAKENHI Grants No.~JP15H05888, No.~JP17H06359, No.~JP16K17707, No.~JP18H04355,
and the MEXT-Supported Program for the Strategic Research Foundation at Private Universities, 2014-2018 (S1411024).
\end{acknowledgments}

%

%-------------------------------------------------------------------%

%-------------------------------------------------------------------%

%---------   References   ---------%

\end{document}